\documentclass[sigconf]{acmart}

\AtBeginDocument{%
  }

\setcopyright{acmlicensed}
\copyrightyear{2018}
\acmYear{2018}
\acmDOI{XXXXXXX.XXXXXXX}
\acmConference[Conference acronym 'XX]{Make sure to enter the correct
  conference title from your rights confirmation email}{June 03--05,
  2018}{Woodstock, NY}
\acmISBN{978-1-4503-XXXX-X/2018/06}

\usepackage{wasysym}
\usepackage{textcomp}
\usepackage{xcolor}
\usepackage{url}
\usepackage{hyperref}
\usepackage{utfsym}
\usepackage{enumitem}
\usepackage{subcaption}
\usepackage{graphicx}
\usepackage{ulem}
\usepackage{multirow}
\usepackage{listings}
\usepackage{color}
\definecolor{pblue}{rgb}{0.13,0.13,1}
\definecolor{pgreen}{rgb}{0,0.5,0}
\definecolor{pred}{rgb}{0.9,0,0}
\definecolor{pgrey}{rgb}{0.46,0.45,0.48}
\usepackage{url}
\usepackage{caption}
\usepackage{bigstrut,bigdelim}
\usepackage{float}

\usepackage{multirow,multicol}
\usepackage[most]{tcolorbox}
\usepackage{enumitem}
\usepackage{colortbl}
\definecolor{mygray}{gray}{.9}
\usepackage{float}

\usepackage{xspace}
\usepackage{tikz}
\usepackage{pgfplots}
\usepackage{colortbl}
\pgfplotsset{compat=1.3}

\usepackage{circledsteps}

\usepackage{xcolor}
\usepackage{enumitem}
\usepackage{hyperref}
\usepackage{algorithm}
\usepackage{algpseudocode}
\usepackage{graphicx}
\usepackage{array} 
\usepackage{multirow}
\usepackage[table]{xcolor}
\usepackage{colortbl}
\usepackage{threeparttable}
\usepackage[most]{tcolorbox}
\usepackage{listings}
\usepackage{lipsum}
\usepackage{breqn} 
\tcbuselibrary{listingsutf8}
\usepackage{nameref}
\tcbuselibrary{listings,breakable}

\usepackage{multirow}
\usepackage[most]{tcolorbox}
\usepackage{graphicx}

\usepackage{xspace}
\usepackage{enumitem}
\usepackage{colortbl}
\usepackage{pifont}

\makeatletter
\DeclareRobustCommand\onedot{\futurelet\@let@token\@onedot}
\def\@onedot{\ifx\@let@token.\else.\null\fi\xspace}
 
\def\eg{\emph{e.g}\onedot} 
\def\ie{\emph{i.e}\onedot} 
 
\def\etc{\emph{etc}\onedot}

\DeclareRobustCommand{\method}{{\sc AutoSW}\xspace}
\DeclareRobustCommand{\textbfmethod}{{\sc \textbf{AutoSW}}\xspace}
\makeatother

\usepackage[framemethod=tikz]{mdframed}
\mdfdefinestyle{customstyle}{
  linecolor=gray!100,
  linewidth=3pt,
  innerleftmargin=3pt,
  topline=false,
  rightline=false,
  bottomline=false,
  leftline=true,
  innerrightmargin=3pt,
  innertopmargin=3pt,
  innerbottommargin=3pt,
  backgroundcolor=gray!15
}
\newenvironment{custommdframed}
  {\begin{mdframed}[style=customstyle]}
  {\end{mdframed}}

\begin{document}


\title{A Viable Paradigm of Software Automation: \\ Iterative End-to-End Automated Software Development}

\author{Jia Li$^{1}$, \ Zhi Jin$^{1,2,*}$, \ Huangzhao Zhang$^{3}$, \ Kechi Zhang$^{2}$, \ Jiaru Qian$^{1}$, \ Tiankuo Zhao$^{1}$ \\
$^1$School of Computer Science, Wuhan University, China \\
$^2$Key Lab of High Confidence Software Technology (PKU), Ministry of Education \\
School of Computer Science, Peking University, China \\
$^3$Independent \\
jia.li@whu.edu.cn, 
zhijin@whu.edu.cn}

\renewcommand{\shortauthors}{Jia Li, Zhi Jin, Huangzhao Zhang, Kechi Zhang et al.}

\begin{abstract}

Software development automation is a long-term goal in software engineering. With the development of artificial intelligence (AI), more and more researchers are exploring approaches to software automation. They view AI systems as tools or assistants in software development, still requiring significant human involvement. Another initiative is ``vibe coding'', where AI systems write and repeatedly revise most (or even all) of the code. We foresee these two development paths will converge towards the same destination: AI systems participate in throughout the software development lifecycle, expanding boundaries of full-stack software development. In this paper, we present a vision of an iterative end-to-end automated software development paradigm \method. It operates in an analyze-plan-implement-deliver loop, where AI systems as human partners become first-class actors, translating human intentions expressed in natural language into executable software. We explore a lightweight prototype across the paradigm and initially execute various representative cases. The results indicate that \method can successfully deliver executable software, providing a feasible direction for truly end-to-end automated software development.

\end{abstract}



\received{20 February 2007}
\received[revised]{12 March 2009}
\received[accepted]{5 June 2009}

\maketitle

\renewcommand{\thefootnote}{\fnsymbol{footnote}}
\footnotetext{*Corresponding author.}

\section{Introduction}

\begin{center}
 \textit{Hand over the key of software development to users.}  \\
\hfill --- \textit{Ruqian Lu}
\end{center}

Software is the soul of the digital world and the cornerstone of contemporary society \cite{celebic2022role, halal1993information, de2025hidden}. Currently, breakthroughs in artificial intelligence (AI), especially large language models (LLMs) and agent technology, have provided unprecedented hope for promoting software development \cite{guo2025deepseek, achiam2023gpt, Gemini-2.5-Pro, Claude-Sonnet-4}. A growing number of researchers and industry professionals have explored development approaches that take advantage of the understanding, reasoning, and generation capabilities of LLMs, with the aim of achieving automated software development \cite{zhang2024codeagent, zhu2024hot, li2025large}. 

AI systems increasingly serve as an auxiliary tool throughout the software lifecycle \cite{hassan2024towards, khemka2024toward}, helping tasks ranging from requirements analysis to code generation and bug fixing \cite{jin2024mare, xia2024agentless}.
A typical case is automated code completion \cite{wang2024rlcoder, husein2025large}, where models predict tokens or lines given the code context.
Although the workload is reduced as a result, human developers remain indispensable, as they need to provide supervision and intervention at each stage and during transitions between stages \cite{qiu2025today, mahmood2022software}.

Meanwhile, in the past year, early adopters have been energetically championing the ``vibe coding'' initiative \cite{Windsurf, Cursor, Augment_Code, Claude_Code}, in which LLMs write and iteratively revise most (or even all) of the code, while human beings steer via natural language prompts.
The approach positions human beings as product managers while offloading implementation to LLMs.
However, current practice often exhibits substantial variability and ad-hoc workflows, raising in the produced programs that are highly coupled and raising concerns about maintainability, quality, and comprehensibility.

We foresee these two lines of development converging on the same destination: AI systems participate across the entire software lifecycle, expanding the boundaries of full-stack software development.
From the earliest days of computing, programmers interacted with machines via punched cards and paper tape;
the advent of magnetic disks and interactive terminals reduced manual toil and enabled assembly programming;
later, compilers and higher-level languages displaced assembly for most purposes.
We argue that we are now at a similar inflection point: a new paradigm of software automation is coming into view, in which AI systems as human partners become first-class actors in design, implementation, and maintenance.
Humans play the role of primary drivers and supervisors in software development.
Looking ahead, AI systems are poised to form a new abstraction layer for software development, translating the intent of humans expressed in natural language into executable software.

\begin{figure}[t] 
\centering
\includegraphics[width=0.95\linewidth]{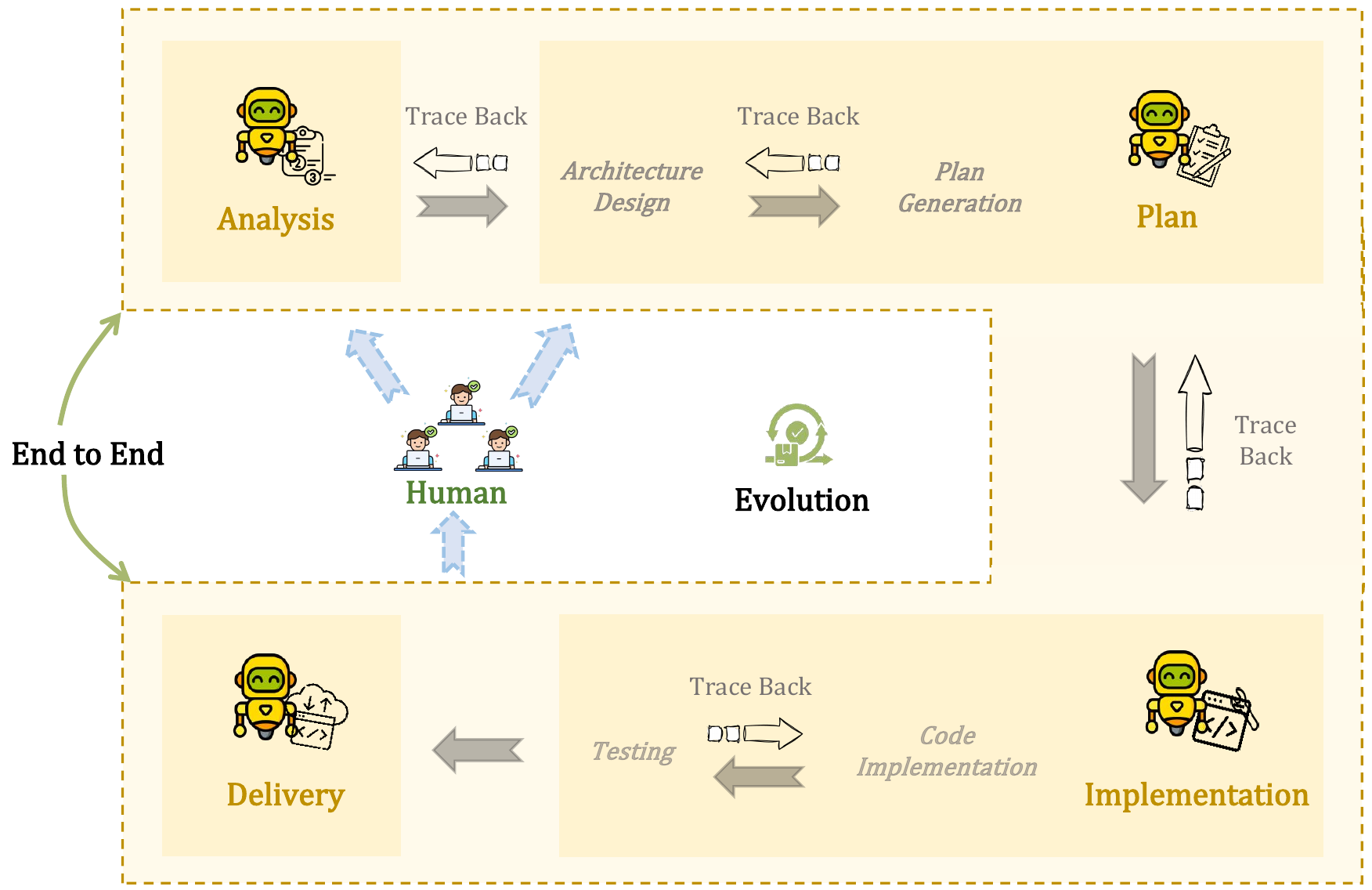}
\caption{Overview of the iterative end-to-end automated
software development paradigm \textbfmethod, operating in an analyze-plan-implement-deliver loop.}
\label{fig: overview}
\end{figure}

In this paper, we present a vision of an iterative end-to-end automated software development paradigm, named \textbfmethod.
This paradigm operates in an analyze-plan-implement-deliver loop (see Figure \ref{fig: overview}), and applies to both greenfield and incremental development.
\ding{182} \textit{Analysis.} AI systems communicate with the human to collect the underlying intent and analyze the requirements gathered. 
\ding{183} \textit{Plan.} Human provides physical constraints in the real-world to AI systems, aligning on the goals of selecting the appropriate technology stack, proposing an architecture design, and eventually producing an implementation plan. 
\ding{184} \textit{Implementation.} Once aligned, AI systems take over and automatically generate source code along with tests, where all functionalities and non-functional requirements are conformed.
\ding{185} \textit{Delivery.} After all tests and verifications have been completed, AI systems produce the deliverable package (code, test reports, deployment status, \etc) and return it to the human.
Through repeated iterations of this loop, the system converges on complex domain-specific software that faithfully conforms to the user’s intent. 

\begin{figure}[t] 
\centering
\includegraphics[width=0.97\linewidth]{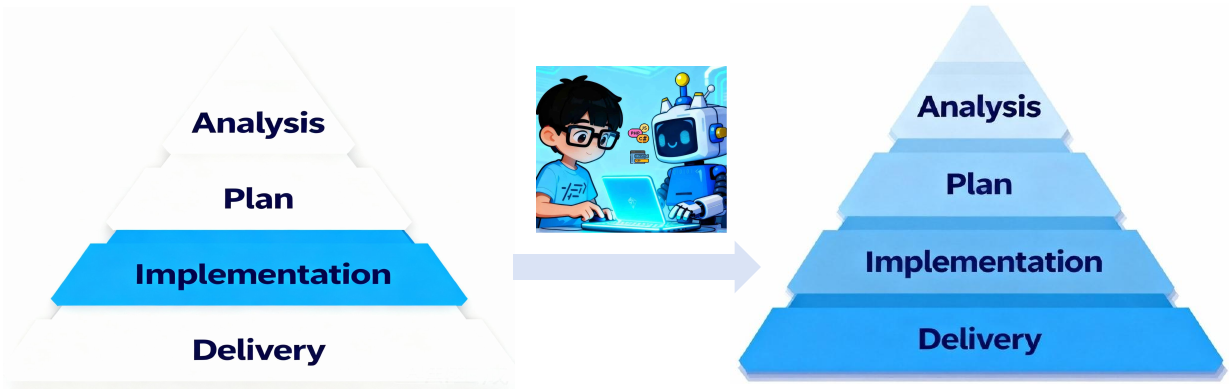}
\caption{Transformation of automated software development paradigm.}
\label{fig: paradigm}
\end{figure}

Figure \ref{fig: paradigm} shows that \method automates the entire software development process, rather than focusing only on code implementation like most existing approaches \cite{zhang2024codeagent, li2023zc,seo2025paper2code, zhang2024codedpo, guo2023longcoder}.
Compared to the prevailing ``vibe coding'' paradigm, \method places a stronger emphasis on software quality. 
It clearly defines the workflow for automated software development and designs a reasonable way to involve human participants.
Through intent alignment, transparent delivery, and traceable automation, \method realizes human requirements in an iterative end-to-end fashion.
Please refer to \S \ref{paradigm} for a detailed introduction to \method, \S \ref{examples} for representative examples in scenarios, and \S \ref{future works} for retrospective and forward-looking discussions of future research directions.

\section{\textbfmethod} \label{paradigm}

This section presents \method, an iterative end-to-end automated software development paradigm based on the orchestral agent. 

\subsection{Overview} \label{overview}

\method is outlined in Figure \ref{fig: overview}, where the AI system as human partner becomes a first-class actor in software automation. Inspired by the classic software development lifecycle, \method includes four stages: \textit{Analysis}, \textit{Plan}, \textit{Implementation}, and \textit{Delivery}. 
In the process, \method maximizes the experience and automation capabilities of the AI system, while focusing on humans to provide constructive information through the conversation with the AI system.

\subsection{Human}

As the primary drivers of software development, humans provide constructive information to the AI system through conversation at two stages: describing requirements at \textit{Analysis} stage, and providing the real-world physical constraints related to the software for \texttt{system design} at \textit{Plan} stage, jointly completing software development from scratch or subsequent evolution. 

In \textit{Analysis} stage, humans describe new or changed requirements to the AI system for requirement analysis, while the AI system proactively initiates conversation by asking questions to progressively achieve requirements clarification. This allows humans to concentrate on describing ``What'' they want, rather than learning ``How'' to provide structured, unambiguous requirements like professional developers, significantly reducing the demand for human expertise.

Considering that system design is fraught with complex trade-offs \cite{gomaa1994software, tavares2010model, cooling2013software}, relying solely on the AI system for design decisions often overlooks real-world physical or business constraints (\eg hardware limitations and specific regulatory requests). \method situates the second human intervention in \texttt{System Design} in \textit{Plan} stage. This intervention focuses humans on providing real-world physical conditions, ensuring the rationality of the selected technology stack and the feasibility of AI-produced architectural decisions.

In \method, humans participate only in indispensable operations and are not required to have knowledge about professional development, maximizing the automation of software development.

\subsection{AI}

The AI partner in \method is an orchestral agent system. Based on requirements and real-world physical constraints provided by humans, the agent system automatically executes the four stages as described in \S \ref{overview}, and finally delivers the deployed software to users. In this section, we elaborate on each stage.

\subsubsection{Analysis} 
The failure in software development usually stems from the ambiguity or contradiction of requirements.
In this stage, the AI system interacts with users to gather the underlying intent and analyze the collected requirements, ultimately producing a standardized requirements document. 
\ding{182} Requirement Elicitation. The agent gradually collects and clarifies user' requirements through active questioning, which mimics the real-world communication pattern between developers and users (\ie concretizing the vague requirements through conversation). This manner supports humans in describing what they want rather than providing structured requirements, allowing laypeople to participate in developing their desired software as well. \ding{183} Requirement Document Generation. Based on the conversation content, a standardized requirements document is generated that serves as a critical bridge between user' intent and engineering practice. The document includes an overall functional description of the software and a serious of user stories. To ensure traceability, each user story has a unique number. \ding{184} Validation. The system then validates requirements from three aspects: correctness, completeness, and consistency, preventing requirement errors from propagating to downstream stages, thus saving significant repair costs.

\subsubsection{Plan} 
This stage aims to create a plan for writing code, involving \texttt{System Design} and \texttt{Plan Generation}, which is pivotal for software quality and maintainability. 
This is the last stage where humans interact with the AI system in order to provide physical constraints related to the software. Once the plan is generated, \method proceeds to \textit{Implementation} stage.

\noindent\texttt{System Design.} 
Software development, especially intricate software, is inseparable from system design \cite{garlan1995introduction, wan2023software}. 
The AI system performs the following actions: \ding{182} Physical Constraint Acquisition. Obtain real-world physical constraints related to software through active conversation with users, since they are crucial for selecting the technology stack and architecture. \ding{183} System Design. Generate an appropriate technology stack and architectures (\ie 4+1 View Model \cite{kruntchen1995architectural}) based on requirements and physical constraints. \ding{184} Validation. Evaluate the system design generated. If errors are discovered, traceable remediation is performed to ensure consistency in all stages.

\noindent\texttt{Plan Generation.}
Plan can help the agent organize the implementation process and prevent any omission of requirements. The AI system generates a series of executable tasks that satisfy the following requests: \ding{182} Each task must be explicit, enabling programming without further clarification. \ding{183} Each task is associated with one or more user stories, which supports a clear traceability matrix from requirements to tasks. \ding{184} The task list must cover all user stories, ensuring that the software can meet the user' intent. \ding{185} The generated plan is not a random list of tasks,  instead, it strictly adheres to the dependencies and modularization defined in the architecture. This ensures the logic and feasibility of implementation, preventing being highly coupled of produced programs. The system then verifies the plan. If errors are found, not only the plan is changed, but the initial error point is also traced back for modification.

\subsubsection{Implementation}
The AI system takes over the plans and automatically generates source code along with tests. This stage contains \texttt{Code Implementation} and \texttt{Testing}.

\noindent\texttt{Code Implementation.}
The AI system is responsible for implementing the tasks defined in the plan, focusing on one task at a time. When users propose new requirements or occur requirement changes, the system must first update the requirement and plan, instead of directly modifying programs.

\noindent\texttt{Testing.} Testing is a key contributor to software quality \cite{murugesan1994attitude}. The AI system evaluates programs from two perspectives: \ding{182} Functionality Testing. It automatically generates test cases based on user stories and acceptance criteria produced by requirement document, testing whether the software behavior meets users' intent or if any redundant functionalities are introduced. \ding{183} Vulnerability Detection. Our ultimate goal is that non-professional humans can also use \method to build their desired software, where they typically provide function-related requirements but cannot mention quality-related demands (\eg preventing SQL injection and XSS attacks). The agent also checks for vulnerabilities to ensure that the software delivered is robust and secure. Once the test fails, it will automatically trigger the traceability mechanism to correct all related errors until the code passes all test cases and does not have vulnerabilities.

\subsubsection{Delivery}

There is a gap between the tested code and the executable software. \method should deliver usable software to users, especially for laypeople, thus achieving an iterative end-to-end loop from the intent of humans to applicable software. In this stage, the AI system completes a series of deployment actions, such as compilation, packaging, configuring servers, and starting services. After deployment, humans can directly use the software and generate feedback. The feedback is then inputted into the AI system through conversations, triggering a new loop if necessary.
\vspace{-1mm}

\subsection{Key Features}

The AI system serving as a human partner in \method is first-class and participates in the entire software development process, which maximizes the automation abilities of the AI system, while focusing on humans providing constructive information through conversation with them in specific stages. In this paper, \method has three key features.

\begin{itemize}[leftmargin=*]
 \item \textbf{Traceability.} 
    \method supports traceability from requirements to testing, where each task in the plan is associated with user stories and can be further traced to code and test cases. When requirements change or errors occur, \method can precisely locate affected components in all stages. This allows for minimal and exact modifications, rather than redeveloping the software. 

    \item \textbf{Software Evolution.} 
    Inspired by the fact that users usually change their requirements, our \method supports software evolution. Users only need to provide new requirements to \method. It will update the corresponding content across relevant stages and then directly delivers a new version of the software to users.

    \item  \textbf{End-to-End Automated Development.}
    Unlike most existing tools, \method starts with the user's intent and ultimately delivers the deployed software. It eliminates the barrier for software development from requiring development knowledge to ideas, empowering any human being to develop their own software.

\end{itemize}

\section{Case Realization} \label{examples}

\subsection{Experimental Setup}
To evaluate the feasibility of \method, we initially implemented a lightweight prototype. In \method, the orchestral agent is instantiated in the state-of-the-art LLM (\ie GPT-5 \cite{GPT-5}). The LLM has impressive understanding, generation, and reasoning abilities, which can support the agent in developing reliable software. 
We execute four representative and realistic scenarios with \method.
Each case is evaluated by two PhD students majoring in computer science. They  verify whether the outcomes satisfy the users' intent.

\subsection{Case Results}
We present the results of \method and analyze their feasibility.
For input, we provide summarized user requirements and physical constraints for convenient demonstration. 

\subsubsection{Game Development}

\noindent \textbf{User Input.}
\textit{Develop a Mine Sweeper game. It enables players to discover mines on the grid and choose different difficulty levels. During game play, it displays the remaining mine count and time statistics.}

\noindent \textbf{Results.}
\method automatically selected the model-view-controller design pattern, dividing the software into a game state management module, a cell-state tracking module, and a user interaction control module. Simultaneously, it separated the game logic from the user interface (UI) rendering and user input, ultimately delivering a fully functional game with smooth interaction.

\vspace{1mm}
\begin{custommdframed}
\textbf{\textit{Finding 1:}} \method is capable of independently designing and implementing state-driven game, successfully converting complex rules of game into operational prototypes.
\end{custommdframed}

\subsubsection{Management System}

\noindent \textbf{User Input.}
\textit{Develop a digital artifact management system that supports the management of multiple types of data, including artifact information, membership data, and user feedback. The data is presented through a visual interface. 
Meanwhile, it enables users to quickly switch between various data modules through interactive operations, with the aim of meeting the daily management of museums.}

\noindent \textbf{Results.} 
\method identified this scenario as a data-intensive application and designed a data-based layout. It successfully integrated a chart library to support data visualization and implemented dynamic filtering logic to support data module switch, where input changes would trigger real-time updates of the data charts.

\vspace{1mm}
\begin{custommdframed}
\textbf{\textit{Finding 2:}} 
\method can build data-intensive applications. The software supports various data operations such as management, analysis, filtering, and visualization.
\end{custommdframed}

\subsubsection{Personal Assistant}

\noindent \textbf{User Input.}
\textit{Develop a travel assistant. It helps users plan their travel itineraries. The assistant first recommends a list of popular attractions and organizes them into an itinerary. When users select an attraction, it can display related weather, flight, accommodation, and navigation information for the location.}

\noindent \textbf{Results.} 
\method successfully designed a multi-modular application, creating distinct modules for attractions, weather, tickets, accommodation, and other functions. The application utilizes a central manager to handle state synchronization across modules. 
When users selected a location in the attractions module, the manager was updated, which in turn automatically triggered the re-rendering of other related modules.

\vspace{1mm}
\begin{custommdframed}
\textbf{\textit{Finding 3:}} \method is capable of developing multi-module applications. The application can effectively keep the communication between modules and the synchronization of their states.
\end{custommdframed}
\vspace{-1mm}

\subsubsection{Application Service}

\noindent \textbf{User Input.}
\textit{Develop a tax filing system. It supports users to complete their tax payments, which enables users to fill out personal information and then automatically analyze their income and generate tax filing materials. To ensure its reliability, the system provides relevant tax policies, explanations, and a Q\&A module to users.}

\noindent \textbf{Results.} 
\method demonstrates an excellent ability to understand and apply domain-specific knowledge, abstracting complex tax policies into a tax filing system that strictly adheres to tax laws. The software not only presents the final results, but also generates a step-by-step breakdown of the calculation process. Additionally, \method autonomously identify sensitive information and implement client-side encryption to protect user privacy.

\vspace{2mm}
\begin{custommdframed}
\textbf{\textit{Finding 4:}} 
\method is able of converting professional and complex domain knowledge (such as tax policies) into precise algorithms and programs. Furthermore, it can autonomously identify and implement non-functional requirements.
\end{custommdframed}
\vspace{0mm}

These results show that \method can complete automated software development in various scenarios, such as state-driven games, data-intensive management systems, multi-module systems, and domain-specific applications. They provide strong empirical support for \method as a viable paradigm of software automation.

\section{Challenges and Future Directions} \label{future works}

Although the results of \method are feasible, there are still some challenges for practical large-scale applications. We outline several directions for future research.

\textbf{Scalability to Complex Scenarios.}
The current exploration focuses primarily on lightweight prototypes. In the future, we will extend \method to complex software scenarios and evaluate its reliability in large-scale applications.

\textbf{Better Form of Human-AI Interaction.}
In \method, humans interact with AI system by multi-turn conversations. 
We will explore better human-AI interaction methods, such as formulating questionnaires that humans only need to fill out quickly, aiming to make it more convenient for humans to interact with AI systems.

\textbf{Context Management.}
During software development, a large amount of context is generated, such as plans, code, test cases, and conversation histories. When developing complex software, it is possible that the context is too long, leading to forgetting long content, such as lost-in-the-middle \cite{liu2023lost, baker2024lost}. We will explore the appropriate memory modules to effectively manage context.

\textbf{Benchmarking.}
The systematically validating of the new paradigm remains an open problem. We will publish a standardized benchmark and design all-sided metrics to evaluate \method. The benchmark will consist of various software scenarios with varying complexity.

\textbf{Enhancing Reliability and Robustness.} 
We will investigate adversarial approaches to enhance its robustness. 
For example, we will require humans to provide contradictory requirements, asking \method to seek clarification rather than directly generating flawed software. 
Meanwhile, We will adopt a more comprehensive software verification strategy to improve reliability.

\section{Conclusion}

This paper presents a vision of an iterative end-to-end automated software development paradigm \method. 
It operates in an analyze-plan-implement-deliver loop, placing a stronger emphasis on software quality. 
In the loop, the AI system serves as the human partner participating in the entire software development process. 
We implement a lightweight prototype and find that \method can successfully deliver satisfactory software in various scenarios.



\bibliographystyle{ACM-Reference-Format}
\bibliography{main}

@String{Computer = "{IEEE} Computer" }

@String{Springer = "Springer-Verlag" }

@misc{celebic2022role,
  title={The Role of Software Engineering in Society 5.0},
  author={Celebic, Vladana},
  year={2022}
}

@article{halal1993information,
  title={The information technology revolution: Computer hardware, software, and services into the 21st century},
  author={Halal, William E},
  journal={Technological Forecasting and Social Change},
  volume={44},
  number={1},
  pages={69--86},
  year={1993},
  publisher={Elsevier}
}

@inproceedings{de2025hidden,
  title={Hidden Figures in Software Engineering: A Replication Study Exploring Undergraduate Software Students' Awareness of Distinguished Scientists from Underrepresented Groups},
  author={de Souza Santos, Ronnie and Santos, Italo and Santos, Robson and Magalhaes, Cleyton},
  booktitle={2025 IEEE International Conference on Software Analysis, Evolution and Reengineering (SANER)},
  pages={786--796},
  year={2025},
  organization={IEEE}
}

@article{Cursor,
  title={Cursor},
  author={},
  journal={https://cursor.com/cn},
  year={2025}
}

@article{Windsurf,
  title={Windsurf},
  author={},
  journal={https://windsurf.com/},
  year={2025}
}

@article{Augment_Code,
  title={Augment Code},
  author={},
  journal={https://www.augmentcode.com/},
  year={2025}
}

@article{Claude_Code,
  title={Claud Code},
  author={},
  journal={https://www.anthropic.com/claude/code},
  year={2025}
}

@article{GPT-5,
  title={GPT-5},
  author={},
  journal={https://openai.com/index/introducing-gpt-5/},
  year={2024}
}

@book{cooling2013software,
  title={Software design for real-time systems},
  author={Cooling, Jim E},
  year={2013},
  publisher={Springer}
}

@article{gomaa1994software,
  title={Software design methods for the design of large-scale real-time systems},
  author={Gomaa, Hassan},
  journal={Journal of Systems and Software},
  volume={25},
  number={2},
  pages={127--146},
  year={1994},
  publisher={Elsevier}
}

@article{tavares2010model,
  title={Model-driven software synthesis for hard real-time applications with energy constraints},
  author={Tavares, Eduardo and Maciel, P and Dallegrave, Pedro and Silva, Bruno and Falc{\~a}o, Tiago and Nogueira, B and Callou, G and Cunha, P},
  journal={Design Automation for Embedded Systems},
  volume={14},
  number={4},
  pages={327--366},
  year={2010},
  publisher={Springer}
}

@inproceedings{wan2023software,
  title={Software architecture in practice: Challenges and opportunities},
  author={Wan, Zhiyuan and Zhang, Yun and Xia, Xin and Jiang, Yi and Lo, David},
  booktitle={Proceedings of the 31st ACM Joint European Software Engineering Conference and Symposium on the Foundations of Software Engineering},
  pages={1457--1469},
  year={2023}
}

@inproceedings{murugesan1994attitude,
  title={Attitude towards testing: a key contributor to software quality},
  author={Murugesan, S},
  booktitle={Proceedings of 1994 1st International Conference on Software Testing, Reliability and Quality Assurance (STRQA'94)},
  pages={111--115},
  year={1994},
  organization={IEEE}
}

@article{baker2024lost,
  title={Lost in the Middle, and In-Between: Enhancing Language Models' Ability to Reason Over Long Contexts in Multi-Hop QA},
  author={Baker, George Arthur and Raut, Ankush and Shaier, Sagi and Hunter, Lawrence E and von der Wense, Katharina},
  journal={arXiv preprint arXiv:2412.10079},
  year={2024}
}

@article{zhang2024codeagent,
  title={Codeagent: Enhancing code generation with tool-integrated agent systems for real-world repo-level coding challenges},
  author={Zhang, Kechi and Li, Jia and Li, Ge and Shi, Xianjie and Jin, Zhi},
  journal={arXiv preprint arXiv:2401.07339},
  year={2024}
}

@article{garlan1995introduction,
  title={Introduction to the special issue on software architecture},
  author={Garlan, David and Perry, Dewayne E},
  journal={IEEE Trans. Software Eng.},
  volume={21},
  number={4},
  pages={269--274},
  year={1995}
}

@article{xia2024agentless,
  title={Agentless: Demystifying llm-based software engineering agents},
  author={Xia, Chunqiu Steven and Deng, Yinlin and Dunn, Soren and Zhang, Lingming},
  journal={arXiv preprint arXiv:2407.01489},
  year={2024}
}

@inproceedings{li2023zc,
  title={ZC 3: Zero-shot cross-language code clone detection},
  author={Li, Jia and Tao, Chongyang and Jin, Zhi and Liu, Fang and Li, Ge},
  booktitle={2023 38th IEEE/ACM International Conference on Automated Software Engineering (ASE)},
  pages={875--887},
  year={2023},
  organization={IEEE}
}

@article{guo2025deepseek,
  title={Deepseek-r1: Incentivizing reasoning capability in llms via reinforcement learning},
  author={Guo, Daya and Yang, Dejian and Zhang, Haowei and Song, Junxiao and Zhang, Ruoyu and Xu, Runxin and Zhu, Qihao and Ma, Shirong and Wang, Peiyi and Bi, Xiao and others},
  journal={arXiv preprint arXiv:2501.12948},
  year={2025}
}

@article{khemka2024toward,
  title={Toward Effective AI Support for Developers: A survey of desires and concerns.},
  author={Khemka, Mansi and Houck, Brian},
  journal={Communications of the ACM},
  volume={67},
  number={11},
  pages={42--49},
  year={2024},
  publisher={ACM New York, NY, USA}
}

@article{kruntchen1995architectural,
  title={Architectural blueprints--the” 4+ 1” view model of software architecture},
  author={Kruntchen, P},
  journal={IEEE software},
  volume={12},
  number={6},
  pages={42--50},
  year={1995}
}

@article{liu2023lost,
  title={Lost in the middle: How language models use long contexts},
  author={Liu, Nelson F and Lin, Kevin and Hewitt, John and Paranjape, Ashwin and Bevilacqua, Michele and Petroni, Fabio and Liang, Percy},
  journal={arXiv preprint arXiv:2307.03172},
  year={2023}
}

@inproceedings{guo2023longcoder,
  title={Longcoder: A long-range pre-trained language model for code completion},
  author={Guo, Daya and Xu, Canwen and Duan, Nan and Yin, Jian and McAuley, Julian},
  booktitle={International Conference on Machine Learning},
  pages={12098--12107},
  year={2023},
  organization={PMLR}
}

@article{wang2024rlcoder,
  title={Rlcoder: Reinforcement learning for repository-level code completion},
  author={Wang, Yanlin and Wang, Yanli and Guo, Daya and Chen, Jiachi and Zhang, Ruikai and Ma, Yuchi and Zheng, Zibin},
  journal={arXiv preprint arXiv:2407.19487},
  year={2024}
}

@article{qiu2025today,
  title={From today’s code to tomorrow’s symphony: The AI transformation of developer’s routine by 2030},
  author={Qiu, Ketai and Puccinelli, Niccol{\`o} and Ciniselli, Matteo and Di Grazia, Luca},
  journal={ACM Transactions on Software Engineering and Methodology},
  volume={34},
  number={5},
  pages={1--17},
  year={2025},
  publisher={ACM New York, NY}
}

@article{mahmood2022software,
  title={Software effort estimation accuracy prediction of machine learning techniques: A systematic performance evaluation},
  author={Mahmood, Yasir and Kama, Nazri and Azmi, Azri and Khan, Ahmad Salman and Ali, Mazlan},
  journal={Software: Practice and experience},
  volume={52},
  number={1},
  pages={39--65},
  year={2022},
  publisher={Wiley Online Library}
}

@article{seo2025paper2code,
  title={Paper2code: Automating code generation from scientific papers in machine learning},
  author={Seo, Minju and Baek, Jinheon and Lee, Seongyun and Hwang, Sung Ju},
  journal={arXiv preprint arXiv:2504.17192},
  year={2025}
}

@article{jin2024mare,
  title={Mare: Multi-agents collaboration framework for requirements engineering},
  author={Jin, Dongming and Jin, Zhi and Chen, Xiaohong and Wang, Chunhui},
  journal={arXiv preprint arXiv:2405.03256},
  year={2024}
}

@article{hassan2024towards,
  title={Towards AI-native software engineering (SE 3.0): A vision and a challenge roadmap},
  author={Hassan, Ahmed E and Oliva, Gustavo A and Lin, Dayi and Chen, Boyuan and Ming, Zhen and others},
  journal={arXiv preprint arXiv:2410.06107},
  year={2024}
}

@article{li2025large,
  title={Large language model-aware in-context learning for code generation},
  author={Li, Jia and Tao, Chongyang and Li♂, Jia and Li, Ge and Jin, Zhi and Zhang, Huangzhao and Fang, Zheng and Liu, Fang},
  journal={ACM Transactions on Software Engineering and Methodology},
  volume={34},
  number={7},
  pages={1--33},
  year={2025},
  publisher={ACM New York, NY}
}

@inproceedings{zhu2024hot,
  title={Hot or cold? adaptive temperature sampling for code generation with large language models},
  author={Zhu, Yuqi and Li, Jia and Li, Ge and Zhao, YunFei and Jin, Zhi and Mei, Hong},
  booktitle={Proceedings of the AAAI Conference on Artificial Intelligence},
  volume={38},
  number={1},
  pages={437--445},
  year={2024}
}

@article{achiam2023gpt,
  title={Gpt-4 technical report},
  author={Achiam, Josh and Adler, Steven and Agarwal, Sandhini and Ahmad, Lama and Akkaya, Ilge and Aleman, Florencia Leoni and Almeida, Diogo and Altenschmidt, Janko and Altman, Sam and Anadkat, Shyamal and others},
  journal={arXiv preprint arXiv:2303.08774},
  year={2023}
}

@article{Gemini-2.5-Pro,
  title={Gemini-2.5-Pro},
  author={},
  journal={https://cloud.google.com/vertex-ai/generative-ai/docs/models/gemini/2-5-pro},
  year={2025}
}

@article{Claude-Sonnet-4,
  title={Claude Sonnet 4},
  author={},
  journal={https://www.anthropic.com/claude/sonnet},
  year={2025}
}

@article{zhang2024codedpo,
  title={Codedpo: Aligning code models with self generated and verified source code},
  author={Zhang, Kechi and Li, Ge and Dong, Yihong and Xu, Jingjing and Zhang, Jun and Su, Jing and Liu, Yongfei and Jin, Zhi},
  journal={arXiv preprint arXiv:2410.05605},
  year={2024}
}

@article{husein2025large,
  title={Large language models for code completion: A systematic literature review},
  author={Husein, Rasha Ahmad and Aburajouh, Hala and Catal, Cagatay},
  journal={Computer Standards \& Interfaces},
  volume={92},
  pages={103917},
  year={2025},
  publisher={Elsevier}
}

\appendix


\end{document}